\documentclass[fleqn,twoside]{article}
\usepackage{amsmath,bm,braket}
\usepackage{espcrc2}
\usepackage{graphicx}
\usepackage{cite}

\DeclareMathOperator{\Li}{\mathrm{Li}}

\allowdisplaybreaks

\title{Infrared Singularities and Soft Gluon Resummation with Massive Partons}

\author{
  Andrea Ferroglia%
  \address[Mainz]{Institut f\"ur Physik (THEP), Johannes Gutenberg-Universit\"at, D-55099
    Mainz, Germany}%
  \address{Physics Department, New York City College of Technology, 300 Jay Street,
    Brooklyn NY 11201, USA},
  Matthias Neubert%
  \addressmark[Mainz]%
  \address{Institut f\"ur Theoretische Physik, Ruprecht-Karls-Universit\"at Heidelberg,
    Philosophenweg 16, D-69120 Heidelberg, Germany},
  Ben D. Pecjak%
  \addressmark[Mainz]
  and Li Lin Yang%
  \addressmark[Mainz]%
  \thanks{Talk given at \textit{Loops and Legs in Quantum Field Theory 2010}, W\"orlitz,
    Germany, April 25-30, 2010.}
}

\begin{document}

\begin{abstract}
  Infrared divergences of QCD scattering amplitudes can be derived from an anomalous
  dimension matrix, which is also an essential ingredient for the resummation of large
  logarithms due to soft gluon emissions. We report a recent analytical calculation of the
  anomalous dimension matrix with both massless and massive partons at two-loop level,
  which describes the two-loop infrared singularities of any scattering amplitudes with an
  arbitrary number of massless and massive partons, and also enables soft gluon
  resummation at next-to-next-to-leading-logarithmic order. As an application, we
  calculate the infrared poles in the $q\bar{q} \to t\bar{t}$ and $gg \to t\bar{t}$
  scattering amplitudes at two-loop order.
\end{abstract}

\maketitle

\section{Introduction}

Due to the complicated self-interactions among the gauge bosons, the structure of infrared
(IR) singularities is a highly non-trivial property of non-abelian gauge theories such as
quantum chromodynamics (QCD). The knowledge of IR divergences in the scattering amplitudes
is essential to the derivation of factorization theorems \cite{Collins:1989gx}. And while
factorization theorems guarantee the absence of IR divergences in sufficiently inclusive
observables, in many cases large Sudakov logarithms remain after this cancellation. A
detailed control over the structure of IR poles in the virtual corrections to scattering
amplitudes is a prerequisite for the resummation of these logarithms beyond the leading
order \cite{Sterman:1986aj, Catani:1989ne}.

For scattering amplitudes involving only massless partons, Catani was the first to predict
the singularities at two-loop order apart from the single pole \cite{Catani:1998bh}, whose
general form was only understood much later in \cite{Sterman:2002qn, Aybat:2006wq,
  Aybat:2006mz}. In recent works \cite{Becher:2009cu, Becher:2009qa}, it was shown that
the IR singularities of on-shell amplitudes in massless QCD can be derived from the
ultraviolet (UV) poles of operator matrix elements in soft-collinear effective theory
(SCET) \cite{Bauer:2000yr, Bauer:2001yt, Beneke:2002ph}. They can be subtracted by means
of a multiplicative renormalization factor, whose structure is constrained by the
renormalization group. The corresponding anomalous-dimension matrix is constrained by
soft-collinear factorization, the non-abelian exponentiation theorem, and the behavior of
scattering amplitudes in two-parton collinear limits. With these constraints, the results
of \cite{Catani:1998bh, Sterman:2002qn, Aybat:2006wq, Aybat:2006mz} can be easily
understood, and it was further proposed that the simplicity of the anomalous-dimension
matrix holds not only at one- and two-loop order, but may in fact be an exact result of
perturbation theory. This possibility was raised independently in \cite{Gardi:2009qi},
while potential new structures at three-loop were investigated in \cite{Dixon:2009ur}.

It is relevant for many physical applications to generalize these results to the case of
massive partons. The IR singularities of one-loop amplitudes containing massive partons
were obtained some time ago in \cite{Catani:2000ef}, but until very recently little was
known about higher-loop results. In the effective theory language used in
\cite{Becher:2009cu, Becher:2009qa}, one needs heavy-quark effective theory (HQET) (see,
e.g., \cite{Neubert:1993mb} and references therein) besides SCET, and the simplicity of
the anomalous dimension matrix observed in the massless case no longer persists in the
presence of massive partons. While the non-abelian exponentiation theorem still restricts
the allowed color structures, important constraints from soft-collinear factorization and
two-parton collinear limits are lost. In \cite{Becher:2009kw}, the generic form of the
anomalous-dimension matrix are given, with two universal functions $F_1$ and $f_2$
representing three-parton correlations left unspecified.

Very recently, in \cite{Ferroglia:2009ep, Ferroglia:2009ii}, the two functions $F_1$ and
$f_2$ were calculated analytically, and therefore the generic two-loop anomalous-dimension
matrix with arbitrary number of massive and massless partons is now completely determined.
These results provide a general, unified, and analytic description of the IR singularities
arising up to two-loop order in arbitrary scattering amplitudes with massive and massless
external legs in unbroken gauge theories such as QCD and QED. They also form the basis for
the systematic resummation at next-to-next-to-leading logarithmic (NNLL) order of large
Sudakov logarithms affecting the rates for such processes. As a first application, the IR
poles in the two-loop virtual amplitudes for $q\bar{q} \to t\bar{t}$ and $gg \to t\bar{t}$
processes were determined in \cite{Ferroglia:2009ep, Ferroglia:2009ii}. Utilizing these
results, an approximate next-to-next-to-leading order (NNLO) formula and an NNLL
resummation formula for top quark pair production at hadron colliders were later derived
in \cite{Ahrens:2009uz, Ahrens:2010zv} with detailed discussions on numerical results and
phenomenological implications. In this talk, we first briefly review the generic setup in
\cite{Becher:2009cu, Becher:2009qa, Becher:2009kw} and then present the results of
\cite{Ferroglia:2009ep, Ferroglia:2009ii}.

\section{Infrared singularities and anomalous dimension matrix}

We denote by $\Ket{\mathcal{M}_n(\epsilon,\{\underline{p}\},\{\underline{m}\})}$ a
UV-renormalized on-shell $n$-parton scattering amplitude with IR singularities regularized
in $d=4-2\epsilon$ dimension. Here $\{\underline{p}\} = \{p_1,\ldots,p_n\}$ and
$\{\underline{m}\} = \{m_1,\ldots,m_n\}$ denote the momenta and masses of the external
partons. The amplitude is a function of the Lorentz invariants $s_{ij} \equiv 2\sigma_{ij}
p_i \cdot p_j + i0$ and $p_i^2=m_i^2$, where the sign factor $\sigma_{ij}=+1$ if the
momenta $p_i$ and $p_j$ are both incoming or outgoing, and $\sigma_{ij}=-1$ otherwise. For
massive partons we define four velocities $v_i=p_i/m_i$ with $v_i^2=1$. We further define
the recoil variables $w_{ij} \equiv -\sigma_{ij} v_i \cdot v_j - i0$. We use the
color-space formalism in \cite{Catani:1996jh} extended to the case with Wilson
coefficients and effective operators, which has been explained in detail in
\cite{Ferroglia:2009ii, Ahrens:2010zv} and will not be repeated here.

It was shown in \cite{Becher:2009cu, Becher:2009qa, Becher:2009kw} that the IR poles of
such amplitudes can be removed by a multiplicative renormalization factor
$\bm{Z}^{-1}(\epsilon,\{\underline{p}\},\{\underline{m}\},\mu)$, which is the same one
used to renormalize the UV divergences of the operator matrix elements in the effective
theory. The renormalization factor satisfied the renormalization group equation (RGE)
\begin{align}
  \label{eq:rge}
  \bm{Z}^{-1} \frac{d}{d\ln\mu} \bm{Z}(\epsilon,\{\underline{p}\},\{\underline{m}\},\mu) =
  -\bm{\Gamma}(\{\underline{p}\},\{\underline{m}\},\mu) \, .
\end{align}
In the case of only massless partons, the anomalous dimension matrix $\bm{\Gamma}$ has the
simple form \cite{Becher:2009cu, Becher:2009qa}
\begin{align*}
  \bm{\Gamma} = \sum_{(i,j)} \frac{\bm{T}_i \cdot \bm{T}_j}{2} \,
  \gamma_{\text{cusp}}(\alpha_s) \ln\frac{\mu^2}{-s_{ij}} + \sum_i \gamma^i(\alpha_s) \, ,
\end{align*}
where the absence of three-parton correlation terms was explicitly demonstrated in
\cite{Aybat:2006wq, Aybat:2006mz}. In \cite{Becher:2009cu, Becher:2009qa, Gardi:2009qi},
it was conjectured that this result may hold to all orders of perturbation theory. On the
other hand, when massive partons are involved in the scattering process, then starting at
two-loop order, correlations involving more than two partons appear. At two-loop order,
the general structure of the anomalous dimension matrix is \cite{Becher:2009kw}
\begin{align}
  &\bm{\Gamma} = \sum_{(i,j)} \frac{\bm{T}_i \cdot \bm{T}_j}{2} \,
  \gamma_{\text{cusp}}(\alpha_s) \ln\frac{\mu^2}{-s_{ij}} + \sum_i \gamma^i(\alpha_s)
  \nonumber
  \\
  &- \sum_{(I,J)} \frac{\bm{T}_I \cdot \bm{T}_J}{2} \,
  \gamma_{\text{cusp}}(\beta_{IJ},\alpha_s) + \sum_I \gamma^I(\alpha_s) \nonumber
  \\
  &+ \sum_{I,j} \bm{T}_I \cdot \bm{T}_j \, \gamma_{\text{cusp}}(\alpha_s)
  \ln\frac{m_I\mu}{-s_{Ij}} \nonumber
  \\
  &+ \sum_{(I,J,K)} if^{abc} \, \bm{T}_I^a \bm{T}_J^b \bm{T}_K^c \,
  F_1(\beta_{IJ},\beta_{JK},\beta_{KI}) \nonumber
  \\
  &+ \sum_{(I,J)} \sum_k if^{abc} \, \bm{T}_I^a \bm{T}_J^b \bm{T}_k^c \, f_2 \left(
    \beta_{IJ}, \ln\frac{-\sigma_{Jk} \, v_J \cdot p_k}{-\sigma_{Ik} \, v_I \cdot p_k}
  \right) \nonumber
  \\
  &+ \mathcal{O}(\alpha_s^3) \, ,
\end{align}
where the lower case indices $i$, $j$, $k$ denote massless partons and the upper case
indices $I$, $J$, $K$ denote massive partons in the external states. The cusp angles
$\beta_{IJ}$ are defined via $\cosh\beta_{IJ}=w_{IJ}$.

The coefficient functions $\gamma_{\text{cusp}}(\alpha_s)$, $\gamma^i(\alpha_s)$ (for $i =
q, g$) have been determined to three-loop order in \cite{Becher:2009qa} by considering the
case of the massless quark and gluon form factors \cite{Moch:2005tm}. Similarly, the
coefficients $\gamma^I(\alpha_s)$ for massive quarks and color-octet partons such as
gluinos have been extracted at two-loop order in \cite{Becher:2009kw} using the known
two-loop anomalous dimension of heavy-light currents in SCET \cite{Neubert:2004dd}. In
addition, the velocity-dependent function $\gamma_{\text{cusp}}(\beta,\alpha_s)$ has been
derived from the known two-loop anomalous dimension of a current composed of two heavy
quarks moving at different velocity \cite{Korchemsky:1987wg, Kidonakis:2009ev}. The most
difficult ones to compute are the two functions $F_1$ and $f_2$, which start at two-loop
order and involve correlations among three partons. They have been derived in closed
analytic form in \cite{Ferroglia:2009ep, Ferroglia:2009ii}, and were confirmed in the
non-physical region by a numerical result in \cite{Mitov:2009sv, Mitov:2010xw}.

It is worth noting here that the anomalous dimension matrix $\bm{\Gamma}$ not only
determines the IR divergences in the amplitudes, but also enters the renormalization group
evolution of the hard functions in the factorization formulas, and therefore serves as the
basis for predicting and resumming logarithmic enhanced terms to all orders in $\alpha_s$.

\section{The functions $F_1$ and $f_2$}

\begin{figure}[t!]
  \begin{center}
    \includegraphics[width=0.32\columnwidth]{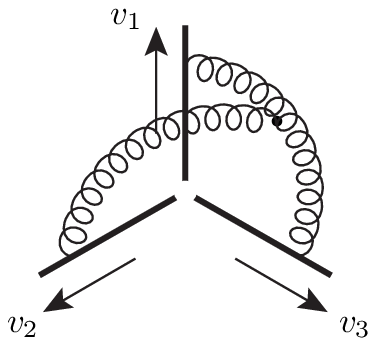}
    \includegraphics[width=0.63\columnwidth]{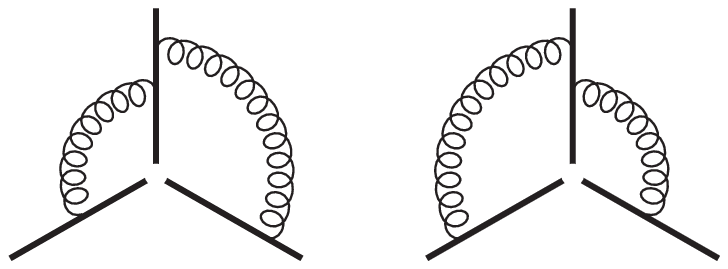}
    \includegraphics[width=0.63\columnwidth]{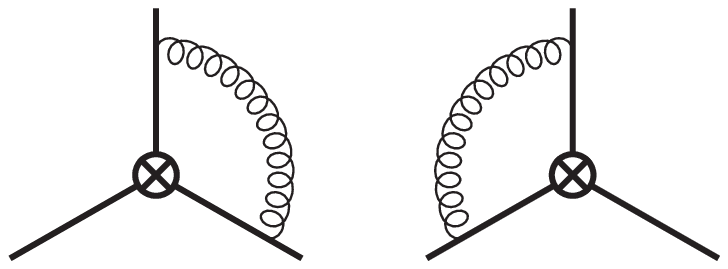}
    \caption{\label{fig:dia} Two-loop Feynman graphs (top row) and one-loop counterterm
      diagrams (bottom row) contributing to the two-loop coefficient of the
      renormalization factor $\bm{Z}_s$.}
  \end{center}
\end{figure}

The calculation of $F_1$ and $f_2$ involves the evaluation of the Feynman diagrams in
Figure~\ref{fig:dia} in the soft limit. More formally, we evaluate the two-loop vacuum
matrix element of the operator $\bm{O}_s=\bm{S}_{v_1}\bm{S}_{v_2}\bm{S}_{v_3}$, which
consists of three soft Wilson lines along the directions of the velocities of the three
partons, without imposing color conservation. For $F_1$, all the three directions are
time-like; while for $f_2$, one of them become light-like. We will describe the
calculation of $F_1$ and then obtain $f_2$ from a limiting procedure.

The operator $\bm{O}_s$ is renormalized multiplicatively so that $\bm{O}_s\bm{Z}_s$ is UV
finite. The anomalous dimension of this operator, which equals $6if^{abc} \,
\bm{T}_1^a\bm{T}_2^b\bm{T}_3^c \, F_1(\beta_{12},\beta_{23},\beta_{31})$, can then be
obtained from $\bm{Z}_s$ by using Eq.~(\ref{eq:rge}). To determine $\bm{Z}_s$, we need to
separate the UV and IR divergences in the amplitude, for that we introduce an offshellness
for the heavy quarks to regulate the IR divergences, while the UV divergences are
regulated dimensionally.

The planar and counter-term diagrams in Figure~\ref{fig:dia} can be evaluated with
standard techniques. The contribution to $F_1$ from those diagrams reads
\begin{multline*}
  F_1^{(2)\,\text{planar+CT}} = \frac{4}{3} \sum_{I,J,K} \epsilon_{IJK} \, \beta_{KI}
  \coth \beta_{KI}
  \\
  \times \coth \beta_{IJ} \Bigg[ \beta_{IJ}^2 + 2\beta_{IJ} \ln(1-e^{-2\beta_{IJ}})
  \\
  - \Li_2(e^{-2\beta_{IJ}}) + \frac{\pi^2}{6} \Bigg] \, .
\end{multline*}
The diagram involving the triple-gluon vertex is the most technically challenging one. We
have computed this diagram using a Mellin-Barnes representation with the result
\cite{Ferroglia:2009ii}
\begin{align*}
  F_1^{(2)\,\text{non-planar}} = -\frac{4}{3} \sum_{I,J,K} \epsilon_{IJK} \beta_{IJ}^2
  \beta_{KI} \coth\beta_{KI} \, .
\end{align*}
For Euclidean velocities, our result agrees numerically with a position-space-based
integral representation derived in \cite{Mitov:2009sv}. Combining all contributions, we
find
\begin{align}
  F_1^{(2)}(\beta_{12},\beta_{23},\beta_{31}) = \frac{4}{3} \sum_{I,J,K} \epsilon_{IJK}
  g(\beta_{IJ}) r(\beta_{KI}) \, ,
\end{align}
where
\begin{align}
  r(\beta) &= \beta \coth\beta \, , \nonumber
  \\
  g(\beta) &= \coth\beta \Bigg[ \beta^2 + 2\beta \ln(1-e^{-2\beta}) \nonumber
  \\
  &\hspace{3em} - \Li_2(e^{-2\beta}) + \frac{\pi^2}{6} \Bigg] - \beta^2 - \frac{\pi^2}{6}
  \, .
\end{align}
The function $f_2$ can be obtained from the above result by writing $w_{23} = -\sigma_{23}
v_2 \cdot p_3/m_3$, $w_{31} = -\sigma_{31} v_1 \cdot p_3/m_3$ and taking the limit $m_3
\to 0$ at fixed $v_I \cdot p_3$. In that way, we obtain
\begin{multline}
  f_2^{(2)} \left( \beta_{12}, \ln\frac{-\sigma_{23} v_2 \cdot p_3}{-\sigma_{13} v_1 \cdot
      p_3} \right)
  \\
  = -4g(\beta_{12}) \, \ln\frac{-\sigma_{23} v_2 \cdot p_3}{-\sigma_{13} v_1 \cdot p_3} \,
  .
\end{multline}
Whether a factorization of the three-parton terms into two functions depending on only a
single cusp angle persists at higher orders in $\alpha_s$ is an open question, but our
results certainly suggest a hidden structure, which seems worthy of further exploration.

It is interesting to investigate the behavior of the two functions $F_1$ and $f_2$ in the
limit where two velocities coincide. This is relevant, e.g., for the situation when two
heavy quarks are produced nearly at rest. Contrary to naive expectations based on
antisymmetry, the two functions don't vanish in this limit. Instead, they tend to a finite
value due to the presence of Coulomb singularities. On the other hand, in the limit when
all the parton masses are small, both functions vanish like $(m_Im_J)^2/s_{IJ}$, in
accordance with a factorization theorem proposed in \cite{Mitov:2006xs, Becher:2007cu}.

\section{Top quark pair production}

Top quark pair production at hadron colliders is an important process in the standard
model. It is also the simplest example which involves two heavy quarks where $F_1$ and
$f_2$ can contribute. Therefore, as a first application, it is interesting to use our
results in this case. In \cite{Ferroglia:2009ep, Ferroglia:2009ii} we have worked out the
relevant anomalous dimension matrices for both $q\bar{q} \to t\bar{t}$ and $gg \to
t\bar{t}$ partonic processes, and derived the IR poles in the two-loop amplitudes. For the
$q\bar{q}$ channel, our results agree with the numerical ones from \cite{Czakon:2008zk},
with the analytic results for some of the color coefficients given in
\cite{Bonciani:2008az, Bonciani:2009nb}, and with the results in the small-mass limit from
\cite{Czakon:2007ej}. For the $gg$ channel, results in the literature are available only
in the small-mass limit \cite{Czakon:2007wk}, and we have checked the agreement of our
exact results with this limiting case. Later at the RADCOR 2009 conference, M.~Czakon
claimed that they had confirmed our results in the $gg$ channel \cite{Czakon:2010rk}.

Given the knowledge of the anomalous dimension matrices, it is possible to derive
approximate formulas beyond next-to-leading order for various differential cross sections
as well as the total cross section for this process. This was done for the invariant mass
distribution in \cite{Ferroglia:2009ep} where all the threshold enhanced terms at NNLO
were exactly predicted. In \cite{Ferroglia:2009ii}, these threshold enhanced terms were
resummed to all orders in $\alpha_s$ at NNLL accuracy, where also the total cross section
was computed by integrating over the invariant mass. Details about these two works are
discussed by A.~Ferroglia in this proceeding.

\section{Conclusions}

The IR divergences of scattering amplitudes in non-abelian gauge theories can be absorbed
into a multiplicative renormalization factor, whose form is determined by an
anomalous-dimension matrix in color space. For processes with only massless partons in the
external states, a simple form of the anomalous-dimension matrix has been conjectured.
Although a lot of supporting arguments were given, a rigorous proof is still lacking. When
massive partons are involved, at two-loop order, the anomalous-dimension matrix contains
pieces related to color and momentum correlations among three partons as long as at least
two of them are massive. This information is encoded in two universal functions: $F_1$,
describing correlations among three massive partons, and $f_2$, describing correlations
among two massive and one massless parton. These two functions have been calculated
analytically in \cite{Ferroglia:2009ep, Ferroglia:2009ii}, and therefore the IR
divergences of any two-loop scattering amplitude with arbitrary number of massive and
massless external particles are completely understood. These results also provide the
basis for a systematic resummation of Sudakov logarithms at NNLL order, and are relevant
to a large class of interesting hadron collider processes and thus have important
implications for precision measurements at the LHC. As first applications, we worked out
the IR poles in the two-loop amplitudes for $q\bar{q} \to t\bar{t}$ and $gg \to t\bar{t}$
processes, derived an approximate NNLO formula for the invariant mass distribution as well
as total cross section for top quark pair production at hadron colliders, and also
resummed the threshold enhanced terms to all orders in $\alpha_s$ at NNLL accuracy.

\end{document}